\documentclass[12pt,a4paper]{article}

\usepackage[T1]{fontenc} 
\usepackage{amsmath}
\usepackage{amsfonts}
\usepackage{amssymb}
\usepackage{graphicx}
\usepackage{subfig}
\usepackage{hyperref}
\newcommand{\HH}{{\cal H}}

\newcommand{\beq}{\begin{equation}}
\newcommand{\eeq}{\end{equation}}
\newcommand{\bea}{\begin{eqnarray}}
\newcommand{\eea}{\end{eqnarray}}
\def\be{\begin{equation}}
\def\ee{\end{equation}}

\def\beq{\begin{equation}}
\def\eeq{\end{equation}}

\newcommand{\rmd}{\textrm{d}}

\newcommand{\pp}{p}
\newcommand{\yy}{y}
\newcommand{\hh}{D_{v}}

\begin{document}
\def\thefootnote{\fnsymbol{footnote}}

\begin{center}
\Large{\textbf{Single-Field Consistency Relations \\of Large Scale Structure}} \\[0.5cm]
 
\large{Paolo Creminelli$^{\rm a}$, Jorge Nore\~na$^{\rm b}$, Marko Simonovi\'c$^{\rm c, \rm d}$ and Filippo Vernizzi$^{\rm e, \rm f}$}
\\[0.5cm]

\small{
\textit{$^{\rm a}$ Abdus Salam International Centre for Theoretical Physics\\ Strada Costiera 11, 34151, Trieste, Italy}}

\vspace{.2cm}

\small{
\textit{$^{\rm b}$ Institut de Ci\`encies del Cosmos (ICC), \\
Universitat de Barcelona, Mart\'i i Franqu\`es 1, E08028- Spain}}

\vspace{.2cm}

\small{
\textit{$^{\rm c}$ SISSA, via Bonomea 265, 34136, Trieste, Italy}}

\vspace{.2cm}

\small{
\textit{$^{\rm d}$ Istituto Nazionale di Fisica Nucleare, Sezione di Trieste, I-34136, Trieste, Italy}}

\vspace{.2cm}

\small{
\textit{$^{\rm e}$ CEA, Institut de Physique Th\'eorique, F-91191 Gif-sur-Yvette c\'edex, France}}

\vspace{.2cm}

\small{
\textit{$^{\rm f}$ CNRS, Unit\'e de recherche associ\'ee-2306, F-91191 Gif-sur-Yvette c\'edex, France}}

\vspace{.2cm}

\end{center}

\vspace{.8cm}

\hrule \vspace{0.3cm}
\noindent \small{\textbf{Abstract}\\ 
We derive consistency relations for the late universe (CDM and $\Lambda$CDM): relations between an $n$-point function of the density contrast $\delta$ and an $(n+1)$-point function in the limit in which one of the $(n+1)$ momenta becomes much smaller than the others. These are based on the observation that a long mode, in single-field models of inflation, reduces to a diffeomorphism since its freezing during inflation all the way until the late universe, even when the long mode is inside the horizon (but out of the sound horizon). These results are derived in Newtonian gauge, at first and second order in the small momentum $q$ of the long mode and they are valid non-perturbatively in the short-scale $\delta$. In the non-relativistic limit our results match with \cite{Kehagias:2013yd,Peloso:2013zw}. These relations are a consequence of diffeomorphism invariance; they are not satisfied in the presence of extra degrees of freedom during inflation or violation of the Equivalence Principle (extra forces) in the late universe.}
\\
\noindent
\hrule
\def\thefootnote{\arabic{footnote}}
\setcounter{footnote}{0}

\newpage

\section{Introduction}

The beautiful results of the  Planck satellite nominal mission \cite{Ade:2013zuv} have disclosed us a very consistent picture of the universe, that firmly confirms the standard cosmological scenario, i.e.~$\Lambda$CDM. This is certainly very comforting. However, the absence of primordial non-Gaussianity in the data \cite{Ade:2013ydc} is somewhat disappointing because the CMB spectrum is above all a test of the physics and of the evolution of the universe at and after recombination and, unfortunately, contains little information on its early stages. Even if a substantial amount of  temperature and polarization data is yet to be disclosed and analyzed, it is unlikely that these new data will change the current overall picture.  

On the other hand, future weak gravitational lensing and galaxy surveys, such as for instance Euclid \cite{euclid2} or Big-Boss~\cite{bigboss}, will be wide and deep enough to give access to many more modes than the CMB. Hence, there is hope that more information on the primordial universe can be captured by these surveys. Such large surveys will  explore very long-wavelength modes, comparable to the current Hubble radius. On these scales, general relativistic effects become important and much theoretical work has been recently done to take  them fully into account (see for example \cite{Yoo:2009au,Bonvin:2011bg,Challinor:2011bk,Yoo:2012se} for galaxies and \cite{Bernardeau:2009bm,Bernardeau:2011tc} for cosmic shear).

One of the primary goals of these surveys is to measure primordial non-Gaussianity. Single-field models of inflation\footnote{Here and in the following when we say ``single-field'' we assume a standard attractor dynamics.} satisfy the well-known Maldacena's consistency relation \cite{Maldacena:2002vr,Creminelli:2004yq,Cheung:2007sv} for the 3-point function in the squeezed limit, which can be derived by using that, for purely adiabatic perturbations, the effect of a long mode with momentum $q$ reduces to a diffeomorphism at zeroth order in $q$. This relation has been later generalized at first order in the momentum $q$ and to arbitrary $n$-point functions \cite{Creminelli:2012ed,Hinterbichler:2012nm,Senatore:2012wy,Hinterbichler:2013dpa,Goldberger:2013rsa}. Consistency relations of this type are very important because 
they provide a model-independent   check of  theoretical calculations in single-field models and, more concretely, because if their violation were observed it would clearly rule out single-field models.

Consistency relations can be reformulated in terms of observables, as nicely reviewed in \cite{Pajer:2013ana}. For instance, in the squeezed limit and when one of the three modes is outside the sound horizon at recombination, the CMB bispectrum  is dominated by the angular modulation of the small-scale power spectrum by long wavelength modes at recombination  and can be rewritten analytically in terms of the angular power spectrum \cite{Creminelli:2004pv,Creminelli:2011sq,Bartolo:2011wb}. Indeed, this analytical relation has been successfully used as a consistency check for second-order Boltzmann codes  \cite{Huang:2012ub}. 

Since a gold mine of data on the large scale structure awaits us in the future, it is natural to try to extend these consistency relations to the fluctuations in the dark matter density or  in the  number density of galaxies. And because upcoming surveys will be sensitive to GR effects, it is reasonable to derive them in a fully relativistic framework. These consistency relations may be even more useful than those for the CMB for two reasons. While CMB temperature fluctuations are very small and can be well understood within linear perturbation theory,  our theoretical control of structure formation is limited, on very small scales, by the nonlinearities in 
the density fluctuations and by the very complex baryonic physics involved in galaxy formation. However, one can derive consistency relations for  large scale structure that hold even when the short-scale modes are deep in the nonlinear regime. Therefore, these relations extend beyond perturbative calculations at second or higher order. The second reason is that such consistency relations hold even when the long wavelength mode is well within the Hubble horizon, simply because during matter dominance or in a $\Lambda$CDM universe, the sound horizon is much smaller than Hubble.

In Sec.~\ref{sec:Weinberg} we first consider the construction of an adiabatic mode developed  by Weinberg, generated by a diffeomorphism at  zeroth order in the momentum $q$ \cite{Weinberg:2003sw}. 
We  generalize this construction at {\em first order} in $q$. To parallel Ref.~\cite{Weinberg:2003sw}  we work in Newtonian gauge (see \cite{Creminelli:2012ed} for an analogous calculation in $\zeta$-gauge). This construction applies in any era of the history of the universe, but here we  concentrate to matter dominance and, later on, we extend our results to a $\Lambda$CDM universe. As mentioned above, during these two stages of the universe the sound speed of fluctuations is very  small. This allows to take the long wavelength mode $q$  {\em well inside the Hubble radius today}, $q \gg aH$, even though it must be kept longer than the sound horizon, i.e.~$q \ll aH/c_s$, which determines the scale at which the long mode interacts with the short ones in a way that cannot be accounted for by a simple coordinate change.

The diffeomorphism generating the adiabatic mode introduced above can be used to study 
how short modes of density fluctuations $\delta$ are modulated by the presence of a long mode, see Sec.~\ref{sec:short_modes}. This is analogous to what has been done to explicitly check second-order calculations during the radiation-to-matter transition \cite{Fitzpatrick:2009ci} or during the whole recombination history of the universe \cite{Creminelli:2011sq}. Even though in this construction we only consider short wavelength fluctuations of the gravitational potential $\Phi$ in the  {\em linear} regime, we do not assume that $\delta$ is small so that the derived results are completely nonperturbative in the short-scale density fluctuations. In Sec.~\ref{sec:cons_rel}, we apply the transformation that induces the long mode to correlation functions and derive consistency relations valid in the squeezed limit. 
We stress here that the  relativistic consistency relations that we formulate are not different from those that apply to inflation \cite{Maldacena:2002vr,Creminelli:2004yq,Cheung:2007sv,Creminelli:2012ed,Hinterbichler:2012nm,Senatore:2012wy,Hinterbichler:2013dpa,Goldberger:2013rsa}. Indeed, their  derivation assumes that the long mode that is inside the horizon today, was  outside the Hubble radius during inflation when the short modes freeze. In other words, the relativistic consistency relations of large scale structure are simply a ``continuation'' of the inflationary consistency relation to the late-time universe.

In the Newtonian limit, our consistency relations reduce to those derived in Refs.~\cite{Kehagias:2013yd} (up to a sign) and \cite{Peloso:2013zw}, using purely Newtonian equations of motion and the perfect fluid approximation to describe dark matter \cite{Bernardeau:2001qr}.  In this regime, the squeezed limit of $(n+1)$-point correlation functions is dominated by the modulation of the bulk (long wavelength) velocity on the $n$-point correlators of the short-wavelength fluctuations. This effect was also discussed in \cite{Jain:1995kx,Scoccimarro:1995if,Bernardeau:2011vy,Bernardeau:2012aq}, where it was shown to vanish when short-wavelength modes are computed at equal time, as also recognized in \cite{Blas:2013bpa,Carrasco:2013sva}.\footnote{The effect of bulk motion on equal-time correlation functions does not vanish for multi-components, in the presence of isodensity fluctuations  \cite{Tseliakhovich:2010bj,Bernardeau:2011vy,Bernardeau:2012aq}. See also footnote 4.}
Here we derive consistency relations independently of the particular small scales dynamics and we show that they are more general than what was discussed in these references. Indeed, 
they apply not only to dark matter density but also to halos or galaxy number density (see also \cite{Kehagias:2013yd}), including nonlinear baryonic effects and bias.

The consistency relations are a consequence of the equivalence principle and they are not related to Galilean invariance \cite{Scoccimarro:1995if,Peloso:2013zw}.
The long wavelength bulk velocity is produced by a homogeneous gradient of the gravitational potential $\Phi$ that, indeed, is captured by the adiabatic mode's construction at first order in $q$. By Euler equation, a constant gradient {\em does not} correspond to a Galilean transformation but to a homogeneous acceleration. Indeed, the fact that small scale correlations are insensitive to the long-scale bulk motion can be simply seen as a consequence of the equivalence principle: one can always remove a homogeneous acceleration by going into a free-falling elevator \cite{Carrasco:2013sva}. In particular, the derived relativistic and Newtonian consistency relations do not apply when gravity is modified---as mentioned in \cite{Kehagias:2013yd}---for example in the absence of equivalence between gravitational and inertial mass. We comment on this point in Appendix \ref{sec:Galilean}.

%%%%%%%%%%%%%%%%%%%%%%%%%%%%%%%%%%%%%%%%%%%%
%%%%%%%%%%%%%%%%%%%%%%%%%%%%%%%%%%%%%%%%%%%%
%%%%%%%%%%%%%%%%%%%%%%%%%%%%%%%%%%%%%%%%%%%%
\section{Adiabatic modes {\em \` a la} Weinberg}
\label{sec:Weinberg}
%%%%%%%%%%%%%%%%%%%%%%%%%%%%%%%%%%%%%%%%%%%%
%%%%%%%%%%%%%%%%%%%%%%%%%%%%%%%%%%%%%%%%%%%%
%%%%%%%%%%%%%%%%%%%%%%%%%%%%%%%%%%%%%%%%%%%%

An adiabatic mode reduces on short scales to a simple change of coordinates. This was used by Weinberg \cite{Weinberg:2003sw} to give a general procedure to write adiabatic modes in the limit in which they are space independent. In this section we are going to generalize this construction to include also a homogeneous gradient of the adiabatic mode.
We start from a linearly perturbed metric  in Newtonian gauge and we consider only scalar perturbations. The metric reads
\be
\rmd s^2 = a^2 (\eta) \left[ - (1+ 2 \Phi) \rmd \eta^2 + (1-2 \Psi) \delta_{ij} \rmd x^i \rmd x^j \right] \;, \label{metric_1}
\ee
where $\Phi$ and $\Psi$ are long wavelength modes. We want to consider a coordinate transformation, at zero momentum, that leaves the metric in the same gauge. Since in Newtonian gauge there is only a conformal factor in front of the spatial coordinates, we are forced to consider a dilation and a special conformal transformation of the spatial coordinates plus a (time-dependent) translation
\begin{align}
\label{etatran} \tilde \eta &= \eta + \epsilon(\eta) + \vec x \cdot \vec \xi (\eta) \;, \\
\label{xtran}\tilde x^i &= x^i (1+ \lambda + 2 \vec x \cdot \vec b) - b^i \vec x^2 + \int^\eta \xi^i \rmd \eta\;,
\end{align}
where $\epsilon$ and $\xi^i$ are arbitrary functions of time and $\lambda$ and $b^i$ are arbitrary constants\footnote{In general we could consider the parameters $\lambda$ and $\vec b$ to be time-dependent. On the other hand, to remain in Newtonian gauge, this would require a change of the conformal time including higher power of $x^i$: one can check that this does not give rise to a physical solution.}. The term proportional to $\xi^i$ in the transformation of the conformal time is chosen to compensate the time-dependent translation and keep the gauge condition $g_{0i} = 0$. In the new coordinates the metric takes the form \eqref{metric_1} with potentials given by
\begin{align}
\tilde \Phi &= \Phi - \epsilon' - \vec x \cdot \vec \xi'  - \HH ( \epsilon + \vec x \cdot \vec \xi) \;, \\
\tilde \Psi &= \Psi + \lambda + 2 \vec x \cdot \vec b + \HH (\epsilon+ \vec x \cdot \vec \xi)\;.
\end{align}
Thus, starting from an unperturbed FRW, we generate a long-wavelength mode of the form 
\be
\Phi =  \epsilon' + \vec x \cdot \vec \xi' +\HH ( \epsilon + \vec x \cdot \vec \xi) \;, \qquad \Psi =  - \lambda - 2 \vec x \cdot \vec b - \HH (\epsilon + \vec x \cdot \vec \xi) \;. \label{conds}
\ee

To ensure that this mode is the long-wavelength limit of a physical solution we have to consider the Einstein's equations that vanish in the limit $q\to 0$ \cite{Weinberg:2003sw}. These are  
\be
(\HH' - \HH^2) v = (\Psi' + \HH \Phi)\;, \label{cond_ct_1}
\ee
where $v$ is the velocity potential and 
\be
\Phi = \Psi - 8 \pi G \delta \sigma\;, \label{cond_ct_2}
\ee
where $\delta \sigma$ is the anisotropic stress, defined by writing the spatial part of the stress-energy tensor as $T_{ij} = g_{ij} p + \partial_i \partial_j \delta \sigma$. Using eqs.~\eqref{conds} and the first condition \eqref{cond_ct_1} one finds that 
\be
\label{v}
v = - (\epsilon + \vec x \cdot \vec \xi) \;,
\ee
and that the comoving curvature perturbation $\zeta \equiv -\Psi + \HH v$ corresponds to  
\be
\label{eq:lambdazeta}
\zeta = \lambda + 2 \vec x \cdot \vec b\;.
\ee
Using the above equation the second condition, eq.~\eqref{cond_ct_2}, reads 
\begin{align}
\epsilon'  + 2 \HH \epsilon  &= - \lambda - 8 \pi G \delta \sigma\;, \label{epseq} \\
\vec \xi' + 2 \HH  \vec  \xi &= - 2 \vec b - 8 \pi G \vec \nabla \delta \sigma\;, \label{xieq}
\end{align}
yielding an equation for $\epsilon$ and $\vec \xi$ in terms of $\lambda$, $\vec b$ and $\delta \sigma$,
\be
\label{epsilon_xi_eqs}
\epsilon = - \frac{1}{a^2} \int a^2 (\lambda + 8 \pi G \delta \sigma) \rmd \tilde \eta\;, \qquad \vec \xi = - \frac{1}{a^2} \int a^2 (2 \vec b + 8 \pi G \vec \nabla \delta \sigma) \rmd \tilde \eta\;.
\ee
If we neglect the anisotropic stress we can write the relation between $\Phi$ and $\zeta$ for the long mode
\be
\label{Phizeta}
\Phi_L = - (\lambda + 2 \vec b \cdot \vec x) (D_v' + {\cal H} D_v) = - \zeta_L (D_v' + {\cal H} D_v) \;,
\ee
where $D_v = \frac{1}{a^2}\int a^2 \rmd \eta$ is the growth function of the velocity.

Taking the gradient of eqs.~\eqref{v} and \eqref{eq:lambdazeta} one obtains, respectively,
\be
\vec \xi = -  \vec v \equiv - \vec \nabla v    \;, \qquad  \vec b = \frac12 \vec \nabla \zeta \;.
\ee
Therefore, one can check using eq.~\eqref{cond_ct_2} and the definition of the comoving curvature perturbation  that eqs.~\eqref{epseq}  and \eqref{xieq} respectively reduce to the Euler equation for the bulk velocity potential $v_L = - \epsilon$ and bulk velocity $\vec v_L = - \vec \xi$,
\be
\label{Euler_L}
v_L ' + \HH  v_L = - \Phi_L\;, \qquad \vec v_L ' + \HH \vec v_L = - \vec \nabla \Phi_L\;.
\ee
Hence, the coordinate transformation \eqref{xtran} corresponds to going into a free-falling elevator and the last term on its RHS is simply the displacement field due to the large-scale velocity $\vec \xi$.

The long mode  leaves the small scale dynamics unaffected---so that its effect can always be removed by a suitable coordinate redefinition---as long as it is much smaller than the sound horizon. Indeed,  in a Hubble time interactions cannot propagate to scales longer than the sound horizon scale. Therefore, our construction holds as long as the long wavelength mode $q$ satisfies $c_s q /(aH) \ll 1$, where $c_s$ is the sound horizon of fluctuations.
Note that for adiabatic perturbations, $\zeta$ is  conserved on  scales longer than the sound horizon, i.e.~$\dot \zeta/(H \zeta) \sim {\cal O} (c_s^2 q^2 /(aH)^2)$, which confirms  eq.~\eqref{eq:lambdazeta}. This implies that, for fluctuations with small sound speed, $c_s \ll1$, such as for cold dark matter, the long wavelength mode can be taken to be much inside  the Hubble radius, $q/(aH) \gg 1$, and still its effect can be removed by a change of coordinates.

Indeed, an important example for the rest of the paper is the regime of {\em matter dominance}, where $c_s \ll 1$. In matter dominance
\be
a \propto \eta^2\;, \quad \mathcal H = \frac 2\eta\;,
\ee
and in the absence of anisotropic stress eq.~\eqref{epsilon_xi_eqs} gives  $\epsilon = - \eta \lambda/5$ and $\vec \xi = - 2 \eta \vec b /5$. In matter dominance eq.~\eqref{Phizeta} boils down 
to $\Phi_L = - \frac{3}{5} \zeta_L$ and the coordinate transformation \eqref{etatran} and \eqref{xtran} becomes
\begin{align}
\tilde \eta &= \eta \Big(1+ \frac13 \Phi_L + \frac13 \partial_i \Phi_L x^i \Big)\;,\\
\tilde x^i & = x^i \Big(1-\frac53 \Phi_L \Big) -\frac53 x^i x^j \partial_j \Phi_L + \frac56 x^2 \partial^i \Phi_L + \frac16 \eta^2 \partial^i \Phi_L \;,
\end{align}
where we have used $\Phi = \Psi$ and the expression for the generated long wave $\Phi$, eq.~\eqref{conds}. Here and in the following $\Phi_L$ and $\partial_i \Phi_L$ do not depend on space and denote the first terms in the Taylor expansion around the origin of the long mode. As in matter dominance the speed of sound of fluctuations is very small, the sound horizon is much smaller than the Hubble radius and our construction holds for a long wave-mode which is much smaller than $H^{-1}$.

Before moving on, let us stress that although here and in the following we stick to a small coordinate transformation that generates a long mode at linear order, it is straightforward to generalize the whole construction to a finite transformation and induce a long mode of large amplitude.

%%%%%%%%%%%%%%%%%%%%%%%%%%%%%%%%%%%%%%%%%%%%
%%%%%%%%%%%%%%%%%%%%%%%%%%%%%%%%%%%%%%%%%%%%
%%%%%%%%%%%%%%%%%%%%%%%%%%%%%%%%%%%%%%%%%%%%
\section{Including short modes}
\label{sec:short_modes}
%%%%%%%%%%%%%%%%%%%%%%%%%%%%%%%%%%%%%%%%%%%%
%%%%%%%%%%%%%%%%%%%%%%%%%%%%%%%%%%%%%%%%%%%%
%%%%%%%%%%%%%%%%%%%%%%%%%%%%%%%%%%%%%%%%%%%%

The procedure discussed above to generate a long mode with a change of coordinates works, almost unchanged, when short modes are also present. This allows to quickly derive second order solutions, as we are going to study in this section. 
The metric in the presence of generic short modes reads
\be
\label{metric_S}
\rmd s^2 = a^2 (\eta) \left[ - (1+ 2 \Phi_S) \mathrm d \eta^2 + (1-2 \Psi_S) \delta_{ij} \mathrm dx^i \mathrm dx^j \right]  \;.
\ee
We assume that the metric is at first order in $\Phi_S$ and $\Psi_S$, while we do not assume that the density contrast $\delta$ is small (this implies that, even in matter dominance, $\Phi_S$ and $\Psi_S$ can be in general time-dependent).

We want to consider a coordinate change that adds to the metric above a long adiabatic mode and thus generates the squeezed limit of the second-order metric in Poisson gauge \cite{Bertschinger:1993xt}.
Let us now consider the change of coordinates \eqref{etatran}, with a modification of the $\eta$ transformation,
\be
\tilde \eta = \eta + \epsilon(\eta) + \vec x \cdot \vec \xi (\eta) + \alpha(\vec x, \eta) \;,
\ee
where $\alpha$ will be of order $\Phi_S$ and it is needed to keep the gauge condition in the presence of the short modes. The gauge condition on the spatial part of the metric is automatically satisfied, the one on off-diagonal terms $\partial_i g_{0i} = 0$ required for Poisson gauge \cite{Bertschinger:1993xt} reads
\be
\nabla^2 \alpha = - 2  \partial_i \left[(\Phi_S (\vec x,\eta) + \Psi_S (\vec x,\eta) )\xi^i(\eta) \right] \;,
\ee
which can be always satisfied with a suitable $\alpha$. 

\subsection{The metric at second order}
Let us verify that with this procedure we can generate the correct second order (in $\Phi$ and $\delta$) solution for the metric in matter dominance. In the following subsection we are going to show that this also reproduces the correct second order solution of $\delta$.
Following the notation of \cite{Boubekeur:2008kn}, the second order solution in Poisson gauge in the squeezed limit reads 
\be
\mathrm d s^2 = \ a^2(\eta)\left[ -\left(1-2\Phi_P  \right) \mathrm d\eta^2
+ \ 2\omega_{Pi}^\perp \mathrm dx_i \mathrm d\eta + \left(1+2\Psi_P  \right)\mathrm d \vec x^2 \right] \;,
\ee
where
\begin{align}
\label{eq:secorder}
\Phi_P &= -\frac{3}{5}\zeta_L + \frac{9}{25}(2\zeta_L \zeta_S - 4\partial_i\zeta_L \partial^{-2}\partial_i \zeta_S) + \frac{6}{25a^2H^2} \partial_i \zeta_L \partial_i \zeta_S \;, \\
\Psi_P &= -\frac{3}{5}\zeta_L - \frac{9}{25}\left(2\zeta_L \zeta_S - \frac{8}{3} \partial_i\zeta_L \partial^{-2}\partial_i \zeta_S \right) \;, \\
\omega_{Pi}^\perp &= -\frac{24}{25aH} (\partial_i\zeta_L \zeta_S - \partial_j\zeta_L \partial^{-2} \partial_i\partial_j\zeta_S) \;.
\end{align}
Here $\zeta_{L,S}$ are the initial curvature perturbation for long and short modes respectively.

We can get the same solution starting from the linear metric \eqref{metric_S} (taking into account that at first order in $\delta$ the potentials are equal $\Phi_S = \Psi_S$ and time-independent) and performing the change of coordinates
\be
\begin{split}
\tilde \eta &= \eta \Big(1+ \frac13 \Phi_L + \frac13 \partial_i \Phi_L x^i \Big)- \frac{4}{3}\eta \,\partial_i \Phi_L \partial^{-2}\partial_i\Phi_S\;, \\
\tilde x^i & = x^i \Big(1-\frac53 \Phi_L \Big) -\frac53 x^i x^j \partial_j \Phi_L + \frac56 x^2 \partial^i \Phi_L + \frac16 \eta^2 \partial^i \Phi_L \;.
\label{eq:changec}
\end{split}
\ee
Using the linear relation $\zeta = -\frac53 \Phi$, we get
\be
\begin{split}
\mathrm d s^2 = & \ a^2(\eta)\left[ -\left(1-2\Phi_P + \frac65  \left(-x^i\zeta_L + \frac{1}{2}\vec x^2  \partial^i \zeta_L - x^i x^j \partial_j\zeta_L  \right)\partial_i\zeta_S\right) \mathrm d\eta^2 \right.  \\
& \left. + \ 2\omega_{Pi}^\perp \mathrm dx_i \mathrm d\eta + \left(1+2\Psi_P - \frac65 \left(-x^i\zeta_L + \frac{1}{2}\vec x^2  \partial^i \zeta_L - x^i x^j \partial_j\zeta_L  \right) \partial_i\zeta_S \right)\mathrm d \vec x^2 \right] \;.
\end{split}
\ee
We indeed find the second order expressions \eqref{eq:secorder} above, with some additional terms. One can check that these describe the action of the long-mode coordinate change (dilations and special conformal transformations) on the initial conditions for the short modes. These encode the inflationary consistency conditions and we are going to come back to them later on. 

Although we are checking it in perturbation theory, let us stress that this construction works {\em non-perturbatively} in the short-scale modes: whatever the short-scale dynamics is, we can generate a new solution that includes a long mode by a coordinate transformation. This statement holds also when baryons are taken into account.

\subsection{The squeezed limit of $\delta^{(2)}$}

In this subsection we want to verify, assuming a matter dominated universe, that our construction above allows to find the density contrast $\delta$ at second order in the perturbations in the squeezed limit at zeroth (this was already done in \cite{Fitzpatrick:2009ci}) and first order in the small momentum $q$. In order to do so we use the change of coordinates eq.~\eqref{eq:changec}.
Since $\rho$ is a scalar, the transformation of $\delta$ induced by the change of coordinates is
\be
\delta \rightarrow \delta + \delta \eta \left( \frac{\bar \rho'}{\bar \rho}(1+\delta) + \delta' \right) + \delta x^i \cdot \partial_i\delta \;,
\label{eq:deltachange}
\ee
with $ \bar\rho \propto \eta^{-6}$ during matter dominance. For clarity let us split the terms ${\cal O}(q^0)$, which correspond to a constant $\Phi_L$, from the ones ${\cal O}(q^1)$, for which the long mode is approximated to have a constant gradient.  

{\bf Constant ${\mathbf \Phi_L}$.} In this case we can neglect terms with $\partial_i\Phi_L$ in eq.~\eqref{eq:changec} and follow \cite{Fitzpatrick:2009ci}. In Fourier space eq.~\eqref{eq:deltachange} transforms the short modes as 
\be
\delta_{\vec k} \rightarrow \delta_{\vec k} - 2\Phi_{\vec q}\delta_{\vec k} + \frac{1}{3} \Phi_{\vec q} \eta \delta_{\vec k}' + \frac{5}{3} \Phi_{\vec q} \partial_{k_i}(k_i\delta_{\vec k}) \;.
\label{eq:deltatrans}
\ee
We can write $\delta_{\vec k}$ as $T_k(\eta)\Phi_{\vec k}$, where $\Phi$ describes the first-order (time-independent) potential in the matter dominated phase, whose value is obviously affected by the evolution during radiation dominance. $T_k(\eta)$ describes the linear relation between $\delta$ and $\Phi$ during MD and it is given by
\be
\label{eq:transfer}
T_k(\eta) \equiv -2 - \frac{k^2\eta^2}{6} \;.
\ee
Plugging this back in the transformation of $\delta$, eq.~\eqref{eq:deltatrans}, we get
\be
\delta_{\vec k} \rightarrow \delta_{\vec k} + \Phi_{\vec q}\Phi_{\vec k} \left( - 2T_k + \frac{1}{3} \eta T_k' + \frac{5}{3} k \partial_{k}T_k \right)  + \frac{5}{3} \Phi_{\vec q} T_k \partial_{k_i}(k_i\Phi_{\vec k})   \;.
\label{dilat_ic}
\ee
The last term describes a dilation of the MD initial conditions. For the time being we are interested in the non-linear evolution during MD. We will come back to the issue of initial conditions in the next section when we discuss consistency relations. Dropping this term we get:
\be
\delta_{\vec k} \rightarrow \delta_{\vec k} + \Phi_{\vec q}\Phi_{\vec k} \left(4 - \frac13 k^2\eta^2 \right)    \;.
\ee
By replacing $\vec q = \vec q_1$ and $\vec k = \vec q_1 + \vec q_2$, this expression reproduces the first two terms of the squeezed limit $q_1 \ll q_2$ of the explicit calculation which, using the notation of \cite{Fitzpatrick:2009ci} and including terms of order ${\cal O}(q_1)$ for future reference, reads
\be
\label{eq:Leosqueezed}
\delta_{\vec k}^{(2)} =  2\left[ 72 - 6q_2^2\eta^2 + \frac{\vec q_1\cdot \vec q_2}{q_2^2} \left( {144} + 11 \eta^2 q_2^2 + \frac{1}{2}\eta^4q_2^4 \right) \right] \frac{\Phi_{\vec q_1} \Phi_{\vec q_2}}{36} \, \big[ 1 + {\cal O}(q_1^2/q_2^2) \big]\;.
\ee
To simplify the notation here and in the following a summation $\int \frac{d^3 q_1\; d^3 q_2}{(2 \pi)^3}   \delta(\vec k - \vec q_1 -\vec q_2)$ is implicitly assumed in products of first order terms in Fourier space.

{\bf Gradient of ${\mathbf \Phi_L}$.} When we add a constant gradient to the coordinate transformation \eqref{eq:changec}  the situation becomes a bit more complicated. Let us look at it  term by term by starting from the time redefinitions, i.e.~the first line of eq.~\eqref{eq:changec}. The last term on the right-hand side gives
\be
\label{eq:PhiLPhiS}
\delta_{\vec k} \rightarrow \delta_{\vec k} + \frac{8 \vec q \cdot \vec k}{k^2}\Phi_{\vec q}\, \Phi_{\vec k} \;,
\ee
while the second term, i.e.~the space-dependent transformation, gives
\be
\delta_{\vec k} \rightarrow \delta_{\vec k} + 2 \Phi_{\vec q}\, q_i\partial_{k_i}\delta_{\vec k} - \frac 13 \Phi_{\vec q}\, q_i \eta \partial_{k_i}\delta_{\vec k}'  \;.
\ee
Replacing  $\delta_{\vec k} = T_k(\eta)\Phi_{\vec k}$, this expression can be rewritten as
\be
\label{sd_time}
\delta_{\vec k} \rightarrow \delta_{\vec k} + \frac{\vec q \cdot \vec k}{k} \Phi_{\vec q} \left(2 \partial_k T_k -\frac13 \eta \partial_k T_k' \right) \Phi_{\vec k} +  \Phi_{\vec q}  \left( 2T_k - \frac{1}{3} \eta T_k' \right) \left(q_j\partial_{k_j}\Phi_{\vec k} \right) \;.
\ee
The last term contains a derivative acting on the initial conditions that we take into account below in this section.

Let us now discuss the spatial redefinitions, i.e.~the second line of eq.~\eqref{eq:changec}. The last term on the right-hand side gives
\be
\label{last_space}
\delta_{\vec k} \rightarrow \delta_{\vec k} - \frac 16 \eta^2 \vec q \cdot  \vec k \, \Phi_{\vec q} T_k \Phi_{\vec k} \;.
\ee
The two remaining terms in the transformation of $x^i$ give 
\be
\delta_{\vec k} \rightarrow \delta_{\vec k} - \frac 56 \Phi_{\vec q} q_j \left[ 2 \partial_{k_i}\partial_{k_j}(k_i\delta_{\vec k}) - \partial_{k_i}\partial_{k_i}(k_j\delta_{\vec k})  \right] \;.
\ee
After replacing  $\delta_{\vec k} = T_k(\eta)\Phi_{\vec k}$, this expression can be conveniently rewritten as
\be
\begin{split}
\label{spatial_red}
\delta_{\vec k} \rightarrow  \delta_{\vec k} - \frac 56 \Phi_{\vec q} \bigg\{  \frac{\vec q \cdot \vec k}{k} \left(4  \partial_{k} T_k + k \partial_{k}^2 T_k \right) \Phi_{\vec k} + 2 q_j  k  \partial_{k} T_k \, \partial_{k_j} \Phi_{\vec k}  + q_j  T_k \left[  2 \partial_{k_i} \partial_{k_j}(k_i\Phi_{\vec k}) - \partial_{k_i}\partial_{k_i}(k_j\Phi_{\vec k})  \right]
 \bigg\} \;.
\end{split}
\ee
Similarly to what happened in eq.~\eqref{dilat_ic}, the last two terms  of this expression---the ones proportional to $T_k$---contain derivatives acting only on the initial conditions. We postpone the discussion of these terms to the next section. In the second term one derivative acts on $T_k$ and the other on $\Phi_{\vec k}$. 
Adding this term to the last one on the right-hand side of eq.~\eqref{sd_time} gives
\be
\label{shift_dil}
-  \Phi_{\vec q} \cdot (q_j\partial_{k_j}\Phi_{\vec k}) \left( -2T_k + \frac{1}{3} \eta T_k' + \frac 53 k_i \partial_{k_i}T_k \right) \;.
\ee
Since $\Phi_{\vec k - \vec q} \simeq \Phi_{\vec k} - (q_j\partial_{k_j}\Phi_{\vec k})$, this is just a shift $\vec k \to \vec k - \vec q$ in the momentum dependence of the initial condition in the dilation transformation, eq.~\eqref{dilat_ic}.

Combining all the contributions, i.e.~eqs.~\eqref{dilat_ic}, \eqref{eq:PhiLPhiS}, \eqref{sd_time}, \eqref{last_space} and \eqref{spatial_red}, and dropping derivatives acting on the initial conditions which as already mentioned will be treated in the next section, we finally obtain 
\be
\begin{split}
\delta_{\vec k} \rightarrow \delta_{\vec k} &+   \left( -2T_k + \frac{1}{3} \eta T_k' + \frac 53 k_i \partial_{k_i}T_k \right) \Phi_{\vec q} \Phi_{\vec k-\vec q}  \\
&+ \frac{\vec q \cdot \vec k}{k^2} \left(8 - \frac43 k \partial_k T_k -\frac56 k^2 \partial_k^2 T_k - \frac13 \eta k \partial_k T_k' - \frac16 \eta^2 k^2 T_k \right) \Phi_{\vec q} \Phi_{\vec k} \;.
\end{split}
\ee
We can now plug in the expression for the transfer function, eq.~\eqref{eq:transfer}, and obtain
\be
\delta_{\vec k} \rightarrow \delta_{\vec k}  +  12 (12 - k^2\eta^2) \frac{\Phi_{\vec q} \Phi_{\vec k-\vec q}}{36} + 2 \frac{\vec q \cdot \vec k}{k^2 } \left( 144  + 23  \eta^2 k^2 + \frac{1}{2}\eta^4k^4 \right) \frac{\Phi_{\vec q} \Phi_{\vec k}}{36}  \;.
\ee
For $\vec q = \vec q_1$ and $\vec k = \vec q_1 + \vec q_2$, this expression reproduces (after expanding the second term on the right-hand side) the squeezed limit $q_1 \ll q_2$ of the explicit result \cite{Fitzpatrick:2009ci}, i.e.~eq.~\eqref{eq:Leosqueezed}.

%%%%%%%%%%%%%%%%%%%%%%%%%%%%%%%%%%%%%%%%%%%%
%%%%%%%%%%%%%%%%%%%%%%%%%%%%%%%%%%%%%%%%%%%%
%%%%%%%%%%%%%%%%%%%%%%%%%%%%%%%%%%%%%%%%%%%%
\section{Consistency relations}
\label{sec:cons_rel}
%%%%%%%%%%%%%%%%%%%%%%%%%%%%%%%%%%%%%%%%%%%%
%%%%%%%%%%%%%%%%%%%%%%%%%%%%%%%%%%%%%%%%%%%%
%%%%%%%%%%%%%%%%%%%%%%%%%%%%%%%%%%%%%%%%%%%%

So far we have discussed how one can add a long mode to a short-scale solution and generate in this way a new solution. Ultimately, if we want to make contact with observations, we have to deal with correlation functions. We now want to show how one can use the same arguments to write down consistency relations: a correlation function with $n+1$ perturbations in the limit in which one of the momenta becomes small compared to the others can be written in terms of a correlation function with $n$ perturbations. Schematically we will have
\be
\langle \Phi_{\vec q}(\eta) \delta_{\vec k_1}(\eta_1) \cdots \delta_{\vec k_n}(\eta_n) \rangle'_{q\to 0} = P_\Phi(q) \sum_a \mathcal O_a \langle \delta_{\vec k_1}(\eta_1) \cdots \delta_{\vec k_n}(\eta_n) \rangle' \;,
\ee 
where $\mathcal O_a$ are differential operators that contain derivatives with respect to the $\vec k_i$ and the $\eta_i$, and correspond to the various terms in the change of coordinates. $P_\Phi(q)$ is the primordial power spectrum of $\Phi$. Here and in the following primes on correlation functions indicate that the momentum conserving delta functions have been removed.

Correlation functions involve statistical averages, so we have to take into account not only the effect of the long mode on the short scale dynamics, but also its effect on the initial conditions of the short modes. If we restrict to single-field models of inflation we know that the long mode acts simply as a coordinate transformation also when initial conditions are set up during inflation: this is indeed the gist of the inflationary consistency relations \cite{Maldacena:2002vr,Creminelli:2004yq,Cheung:2007sv,Creminelli:2012ed, Hinterbichler:2012nm,Senatore:2012wy,Hinterbichler:2013dpa,Goldberger:2013rsa}. Therefore we have a unified picture where the long mode {\em always} acts on the short scales as a coordinate transformation, from inflation until the matter and $\Lambda$ dominated phases. This is what allows to write consistency relations. There is however an important caveat to this logic. For the long mode to be traded for a coordinate redefinition we need it to be outside the sound horizon since the moment the short modes freeze during inflation until the observation today. {\em This in particular requires that the long mode does not enter the sound horizon during radiation dominance,}\footnote{
At recombination, baryons and cold dark matter  have different velocities because baryons are tightly coupled to photons and display acoustic oscillations. As discussed in \cite{Tseliakhovich:2010bj,Bernardeau:2011vy,Bernardeau:2012aq}, the long-wavelength relative velocity between baryons and dark matter reduces the formation of early structures on small scales, through a genuinely nonlinear effect. Note, however, that on wavelengths longer than the sound horizon at recombination, the velocities of dark matter and baryons are the same. Thus, in the limit that we consider here of a long mode that has always been outside the sound horizon, this effect is absent.
} otherwise one has to take into account the dynamics of the long mode interacting with the short ones, and this cannot be described in terms of a change of coordinates. Therefore in the following we are going to assume that the long mode entered the Hubble radius during matter dominance. Moreover, the long mode must of course be much longer than the short ones, independently of whether the latter entered the Hubble radius during matter or radiation dominance.

One can start in real space with the statement that the correlation function of the short-scale matter perturbations in the presence of a long mode of the gravitational potential $\Phi_L$ is the same as the correlation function of the short perturbations in the absence of the long mode but in transformed coordinates,
\be
\langle \delta(\vec x_1,\eta_1) \cdots \delta(\vec x_n,\eta_n) | \Phi_L \rangle = \langle \delta(\vec {\tilde x}_1,\tilde \eta_1) \cdots \delta(\vec {\tilde x}_n,\tilde \eta_n) \rangle \;.
\ee 
Let us stress again that this statement deals also with initial conditions and requires a single-field model of inflation: in a multi-field model the effect of the long mode on the initial conditions {\em is not} a coordinate transformation and the equation above will contain additional terms from local non-Gaussianities.
If we now multiply both sides by $\Phi_L$ and take the average over it, we get on the left-hand side the $(n+1)$-point function with the gravitational potential. On the right-hand side we have to consider the effect of the change of coordinates (otherwise the statistical average with $\Phi_L$ gives zero). Using $(\vec x_a,\eta_a)\equiv x_a$, we can write
\be
\langle \Phi_L(x) \delta(x_1) \cdots \delta(x_n) \rangle = \big\langle \Phi_L(x)\; \frac{\partial}{\partial \Phi_L} \langle \delta(x_1) \cdots \delta(x_n) \rangle_{\Phi_L} \cdot \Phi_L(\yy)\big\rangle \;,
\ee 
where $\yy \equiv (\vec \yy,\eta)$ and $\vec \yy$ is an arbitrary point---for example the midpoint for $\vec x_1,\ldots, \vec x_n$---whose choice is immaterial at order $q$. Using eq.~\eqref{eq:deltachange} one has
\be
\label{consist_real}
\langle \Phi_L(x) \delta(x_1) \cdots \delta(x_n) \rangle = \Big\langle \Phi_L(x) \sum_a \big \langle \delta(x_1) \cdots \left[ \delta \eta_a \left( \frac{\bar \rho'}{\bar \rho}(1+\delta) + \delta' \right) + \delta x^i_a \cdot \partial_i\delta \right]_{x_a} \cdots \delta(x_n) \rangle \Big \rangle \;.
\ee 
Now we have to go to Fourier space as we did above and find the operators ${\mathcal O}_a$. All terms in the coordinate transformation are linear in $\Phi_L$ and their contributions can be evaluated separately.

\subsection{Matter dominance}
Let us first focus on the case of a matter-dominated universe and see what \eqref{consist_real} gives in this case.

{\bf Transformation of spatial coordinates}. The time independent part of the first line of eq.~\eqref{eq:changec} describes a dilation and a special conformal transformation of spatial coordinates,
\be
\eta \rightarrow \eta\;, \quad x^i \rightarrow x^i - \frac 53 \Phi_L x^i - \frac 53 \partial_j\Phi_L x^j x^i + \frac{5}{6} \vec x^2 \partial^i \Phi_L \;.
\ee
These have already been studied in the context of inflation \cite{Creminelli:2012ed} and we can follow exactly the same algebra to get
\be
\label{conformal}
\langle \Phi_{\vec q} \, \delta_{\vec k_1}(\eta_1) \cdots \delta_{\vec k_n}(\eta_n) \rangle'_{q\to 0} \supset  \frac 53 P_\Phi(q)\left( 3(n-1) + \sum_a \vec k_{a} \cdot \vec \partial_{k_{a}} + \frac 12 q^iD_i \right) \langle \delta_{\vec k_1}(\eta_1) \cdots \delta_{\vec k_n}(\eta_n) \rangle' \;,
\ee
where $D_i$ is the special conformal transformation written in momentum space
\be
\label{cons:conformal}
\quad q^iD_i \equiv \sum_{a=1}^{n} \Big[ 6\vec{q}\cdot \vec{\partial}_{k_a} - \vec{q}\cdot \vec{k}_a \vec{\partial}_{k_a}^2 +2\vec{k}_a \cdot \vec{\partial}_{k_a} (\vec{q}\cdot \vec{\partial}_{k_a}) \Big] \;.
\ee 
Two comments are here in order. First, notice that this transformation of the $n$-point function contains many physically different contributions: the primordial correlation induced during inflation between the long and the short modes---i.e.~the inflationary conformal consistency relations, the effect of the long mode during radiation dominance and finally during the matter dominated phase. The sum of these effects gives eq.~\eqref{conformal}. The inflationary initial conditions appear only in this term because they do not depend on time and the time-dependent term of the spacial coordinate transformation decays at early times.
The second comment is that the primes on the correlation functions mean that the \emph{same} momentum conservation delta function $(2\pi)^3\delta(\vec q + \sum_i \vec k_i)$ is removed from the correlation functions on both sides: this is non-trivial since the correlation function on the RHS does not contain $q$.  However one gets this result taking into account terms proportional to derivatives of the delta function in momentum space: we refer the reader to \cite{Creminelli:2012ed} for the details of the derivation.

Let us consider next the time-dependent term in the transformation of the spatial coordinates
\be
\eta \rightarrow \eta\;, \quad x^i \rightarrow x^i + \frac{1}{6}\eta^2 \partial^i\Phi_L \;.
\ee
As we have discussed above, this term is the largest on short scales, so that it will dominate the consistency relation when the short modes are deep inside the Hubble radius (which must be the case if the long mode is also inside the horizon). Indeed this contribution can be understood Newtonianly \cite{Kehagias:2013yd, Peloso:2013zw} and it captures a Newtonian non-linearity which becomes very large on short scales, while the other effects that we discuss are purely relativistic.
The variation of the $n$-point correlation function is given by
\be
\delta \langle \delta(x_1) \cdots \delta(x_n) \rangle = \int \frac{\rmd^3\vec \pp}{(2\pi)^3} \frac{\rmd^3\vec k_1}{(2\pi)^3} \cdots\frac{\rmd^3\vec k_n}{(2\pi)^3} \sum_a \left( -\frac 16 \eta_a^2 \, \vec \pp \cdot \vec k_{a} \right) \Phi_{\vec \pp} \, \langle \delta_{\vec k_1} \cdots \delta_{\vec k_n} \rangle e^{i\vec \pp \cdot \vec \yy + i\sum_b \vec k_b \cdot \vec x_b} \;.
\ee
If we multiply by $\Phi_{\vec q}$ and average, on the LHS we are going to get the $(n+1)$-point correlation function. On the RHS the average over $\Phi_{\vec q}$ will give the power spectrum of the long mode. Notice that due to the conservation of momentum, $\vec \pp$ becomes $-\vec q$ and we can finally write
\be
\label{Riotto}
\langle \Phi_{\vec q} \, \delta_{\vec k_1}(\eta_1) \cdots \delta_{\vec k_n}(\eta_n) \rangle'_{q\to 0} \supset  \frac 16 P_\Phi(q) \sum_a \vec q \cdot \vec k_a \eta_a^2 \langle \delta_{\vec k_1}(\eta_1) \cdots \delta_{\vec k_n}(\eta_n) \rangle' \;.
\ee
This generalizes the consistency relation obtained in \cite{Kehagias:2013yd} and \cite{Peloso:2013zw} since it does not assume the long mode to be deep inside the Hubble radius and thus in the Newtonian regime. Notice however that, once written all in terms of $\delta$ such as below (see eq.~\eqref{NRgeneral}), we have an overall sign difference with respect to \cite{Kehagias:2013yd} due a typo in their derivation. Notice also that the RHS of eq.~\eqref{Riotto} vanishes, up to ${\cal O}(q)$, when the short modes are taken at the same time by momentum conservation.

{\bf Transformation of the time coordinate.} 
It is evident from \eqref{eq:deltachange}  that there are two kinds of contributions: one coming from the change of time in the argument of the $\delta$'s and the other from the change of time in the background density $\bar \rho$. We will treat these two separately.
The last term of the first line of \eqref{eq:changec} is suppressed, compared for example to the one on its left, by $\Phi_S$, so it is intrinsically a higher order contribution and we have only to consider its effect on the unperturbed energy density $\bar\rho$ and not on the $\delta$'s.

Therefore, to find the contribution from the change of the argument of density contrast, it is enough to consider
\be
\eta \rightarrow \eta + \frac{1}{3} \Phi_L\eta + \frac{1}{3} x^i\partial_i \Phi_L \eta \;, \quad x^i \rightarrow x^i \;.
\ee
Following \eqref{consist_real}, the variation of the $n$-point function is given by
\be
\begin{split}
\delta \langle \delta(x_1) \cdots  \delta(x_n) \rangle & = \int \frac{\rmd^3\vec \pp}{(2\pi)^3} \frac{\rmd^3\vec k_1}{(2\pi)^3}  \cdots\frac{\rmd^3\vec k_n}{(2\pi)^3} \sum_a \frac 13 \Phi_{\vec \pp} \, \eta_a \partial_{\eta_a}\langle \delta_{\vec k_1} \cdots \delta_{\vec k_n} \rangle (1+ \vec \pp \cdot \vec \partial_{k_{a}}) e^{i\vec \pp\cdot \vec \yy + i\sum_b \vec k_b \cdot \vec x_b}  \\
& = \int \frac{\rmd^3\vec \pp}{(2\pi)^3}  \frac{\rmd^3\vec k_1}{(2\pi)^3}  \cdots \frac{\rmd^3\vec k_n}{(2\pi)^3} \sum_a \left( \frac 13 \Phi_{\vec \pp} \, (1- \vec \pp \cdot \vec  \partial_{k_{a}})\eta_a \partial_{\eta_a} \right) \langle \delta_{\vec k_1} \cdots \delta_{\vec k_n} \rangle e^{i\vec \pp\cdot \vec \yy + i\sum_b \vec k_b \cdot \vec x_b} \;,
\end{split}
\ee
with the same notation as before. Similarly to what happens in the case of special conformal transformations, when the derivatives act on the delta function of momentum conservation they combine with zeroth order term in gradients. In particular
\be
(1- \vec \pp \cdot \vec \partial_{k_{a}}) \delta(\vec k_1 + \cdots \vec k_n) = \delta(-\vec \pp +\vec k_1 + \cdots \vec k_n) \;.
\ee
This allows us to have the same overall delta function on both sides of the consistency relation. Averaging over the long mode $\Phi_{\vec q}$ we find that the contribution to the consistency relation is
\be
\label{time}
\langle \Phi_{\vec q} \, \delta_{\vec k_1}(\eta_1) \cdots \delta_{\vec k_n}(\eta_n) \rangle'_{q\to 0} \supset  \frac 13 P_\Phi(q) \sum_a  \left( 1+ \vec q \cdot \vec \partial_{k_{a}} \right) \eta_a\partial_{\eta_a}\langle \delta_{\vec k_1}(\eta_1) \cdots \delta_{\vec k_n}(\eta_n) \rangle' \;.
\ee

Now we can look at the contributions to the consistency relation coming from the change of the background density  $\bar \rho$. In particular, as here we are only interested in how $\delta$ changes at second-order, we need to consider only two contributions: one coming from the change $\delta \to \delta + \delta \eta ({\bar \rho'}/{\bar \rho}) \delta $ and taking the time coordinate transformation at first order in the long mode, and the other one from considering $\delta \to \delta + \delta \eta ({\bar \rho'}/{\bar \rho}) $ and using the second-order piece of the time coordinate transformation, i.e.~the last term on the first line of eq.~\eqref{eq:changec}. In particular, we can rewrite the latter contribution by using the 00 component of Einstein equations  in matter dominance, $ \partial^2 \Phi_S - 3 \HH^2 \Phi_S = \frac{3 }{2 } \HH^2 \delta$, to replace the short-scale gravitational potential $\Phi_S$ in terms of $\delta$. Finally, putting together these two contributions
we obtain
\be
\delta \to \delta + \frac{\bar \rho'}{\bar \rho} (1+ \delta) \delta\eta\;  = \delta -2(\Phi_L + x^i\partial_i\Phi_L)\delta + 12 \partial_i\Phi_L    \frac{\HH^2 }{ \partial^2 ( \partial^2 - 3 \HH^2)} {\partial_i} \delta \;.
\ee
The contribution to the consistency relation is then
\be 
\label{background}
\langle \Phi_{\vec q} \, \delta_{\vec k_1}\cdots \delta_{\vec k_n} \rangle'_{q\to 0}   \supset    -2 P_\Phi(q) \left( n+ \sum_a \vec q \cdot \vec \partial_{k_{a}} -  \sum_a \frac{4 \vec q\cdot \vec k_a }{k_a^2 (2+k_a^2 \eta_a^2/6)}\right) \langle \delta_{\vec k_1}(\eta_1) \cdots \delta_{\vec k_n}(\eta_n) \rangle'  \;.
\ee

Summing up all contributions from eqs.~\eqref{conformal}, \eqref{Riotto}, \eqref{time},  and \eqref{background} we finally get
\be \label{consistency}
\boxed{
\begin{split}
&\langle \Phi_{\vec q} \, \delta_{\vec k_1}(\eta_1) \cdots \delta_{\vec k_n}(\eta_n) \rangle'_{q\to 0} = P_\Phi(q)\left[ 3n-5 + \frac 13 \sum_a \left(5 \vec k_{a} \cdot \vec \partial_{k_{a}} + \eta_a\partial_{\eta_a} \right)  \right. \\
& \left.  +\frac 16 \sum_a \vec q \cdot \vec k_a \eta_a^2 +\frac 56 q^iD_i  -2 \sum_a \left( 1-\frac 16 \eta_a\partial_{\eta_a} \right) \vec q \cdot \vec \partial_{k_{a}} + \sum_a \frac{8 \vec q\cdot \vec k_a }{k_a^2 (2+k_a^2 \eta_a^2/6)} \right] \langle \delta_{\vec k_1}(\eta_1) \cdots \delta_{\vec k_n}(\eta_n) \rangle'
\;.
\end{split}}
\ee
This is the relativistic consistency relation for large scale structure and the main result of this article. On the RHS, the first line contains terms of order ${\cal O}(q^0)$ while the second terms ${\cal O}(q^1)$.

For $n=2$ we can use this expression to relate the squeezed limit of the 3-point correlation function to the 2-point function.
Indeed, the 2-point function of $\delta$ simply reads
\be
\langle \delta_{\vec k_1} (\eta_1)\delta_{\vec k_2} (\eta_2) \rangle' = \frac{1}{36}P_\Phi(k) (12 + k_1^2\eta_1^2)(12 + k_2^2 \eta_2^2) \;,
\ee
where $k \equiv |\vec k_1 - \vec k_2|/2$ and we stress again that $P_\Phi(k)$ is the power spectrum of $\Phi$ in the matter dominated phase, whose scale dependence is affected by the evolution during radiation dominance.
Plugging this expression on the RHS of  eq.~\eqref{consistency} and expanding for small $q$ using $\vec k_1 \equiv \vec k-  \vec q/2$, $\vec k_2 \equiv - \vec k - \vec q/2$, we obtain
\be
\begin{split}
\langle \Phi_{\vec q} \, \delta_{\vec k_1}(\eta_1) \delta_{\vec k_2} (\eta_2) \rangle'_{q\to 0}  & =  P_\Phi(q)P_\Phi(k)\left(  \frac{-144 + k^4 \eta_1^2 \eta_2^2} {9} + \vec q \cdot \vec k (\eta_1^2 - \eta_2^2) \frac{-168 + k^4 \eta_1^2 \eta_2^2 + 12 k^2 (\eta_1^2 + \eta_2^2)}{216} \right) \\
 &+  \frac53 P_\Phi(q) \frac1{k^3}\frac{d (k^3 P_\Phi(k))}{ d \ln k} \big[ (12 + k^2 \eta_1^2 ) (12 + k^2 \eta_2^2 )  -12 \vec q \cdot \vec k  (\eta_1^2 - \eta_2^2 )\big] \;.
\end{split}
\ee
We can check that this  relation correctly reproduces the 3-point correlation function in the squeezed limit. Indeed, the first line on the RHS of this equation contains the nonlinear evolution in matter dominance. One can verify this by computing the 3-point function using the result of the exact second-order calculation  from Ref.~\cite{Cheung:2007sv} taken in the squeezed limit, i.e.~eq.~\eqref{eq:Leosqueezed}, and expanding it for small $q$.  The second line takes into account the initial condition and directly follows from the Maldacena's calculation of the primordial 3-point correlation function  of $\zeta$ \cite{Maldacena:2002vr}, rewritten in terms of $\Phi$ in matter dominance and also taken in the squeezed limit (the conformal transformation of the initial conditions vanishes when acting on a 2-point function).

Some comments on the consistency relation, eq.~\eqref{consistency}, are in order here:
\begin{enumerate}
\item This equality has been derived at second order in the metric perturbations. Higher order contributions are irrelevant and have been neglected. However, we did not assume at any point that $\delta$ is small. Thus, {\em eq.~\eqref{consistency} is valid deep in the non-linear regime of structure formation.}
\item As already mentioned, this relation is valid in the squeezed limit and deviations from this limit enter as $ {\cal O}(q^2/k^2)$ corrections. To derive it we assumed  that the long mode stays out of the sound horizon during radiation dominance and for this reason we expect it to hold up to $ {\cal O} (q^2/( a H)_{\rm eq}^2)$ corrections. It is valid as long as the long mode is out of the sound horizon; deviations are of order ${\cal O} (c^2_s q^2 /(aH)^2)$.
\item These relations do not assume that the short modes can be treated perturbatively, and as such they are valid also when the short modes are deep in the non-linear regime, even when we take into account the effect of baryons and their bias. Indeed it is important to notice that even if we wrote our relation thinking about $\delta$ as the dark matter density contrast, nothing would change if instead we used $\delta_h$ or $\delta_g$, i.e.~the number density of haloes or of galaxies\footnote{A modification to our formulas will come from the fact the number of haloes or galaxies is not conserved, but evolves in time, so that the effect of time redefinition will be different.}. Since the long mode does not affect the physics at short scale, it will not modify in any way the (linear or non-linear) bias.
\item When the short modes are deep inside the Hubble radius, the first term of the second line of this relation  dominates. This is a Newtonian contribution that describes the effect of the bulk velocity due to the long mode on the short structures. Its effect on equal time correlators cancels due to momentum conservation \cite{Jain:1995kx,Scoccimarro:1995if,Bernardeau:2011vy,Bernardeau:2012aq,Blas:2013bpa,Carrasco:2013sva}. This is the term that enters in the consistency relation in the Newtonian regime discussed in \cite{Kehagias:2013yd} and \cite{Peloso:2013zw} (with the change of sign discussed above). Notice however that our derivation is more general, because it does not start from the dynamical equations describing dark matter only: as we discussed above the relation holds also when baryonic physics is taken into account and it thus includes bias. Notice also that if we are only interested in the (Newtonian) term which dominates when the long mode is inside the Hubble radius, we can disregard the assumption that the long mode has been always outside the sound horizon during the whole cosmological evolution.

\item When we consider the relation between the 3- and 2-point functions, then the contributions of order ${\cal O}(q)$ cancel out when the short modes are evaluated at the same time. Indeed,  the 2-point function does not have any preferred direction and it cannot define a vector that can be contracted with $\vec q$.
\item Delta function of momentum conservation is the same on the two sides of the consistency relation. However, the correlation functions have different number of fields. On the LHS there is no ambiguity, but on the RHS the correlation function has to be evaluated for configuration of momenta that do not form a closed polygon. For this reason, the equation might seem ambiguous. In principle, we can add to the correlator on the RHS any expression that is zero for $\sum \vec k_a=0$, but it is far from obvious that the complicated differential operator will always give zero acting on this term. However, one can show that this is indeed what happens. If we assume that additional term is translationary and rotationally invariant, then the following relation holds
\be
\sum_a  O_a [P_i\mathcal M_i(\vec k_1, \ldots, \vec k_n)]=  \mathcal O(q^2) \;,
\ee
where $\vec P=\sum \vec k_a$, $\mathcal  M$ is an arbitrary (translationary and rotationally invariant) function of momenta and $O_a$ is a differential operator on the RHS of the consistency relation.
\end{enumerate}

%{\bf [Shall we rewrite here the consistency relation \eqref{consistency}  only in terms of $\delta$'s?]}

\subsection{$\Lambda$CDM}

So far we have considered only a matter-dominated universe. However, our construction of the adiabatic mode is valid for an arbitrary background cosmology as long as the long mode is outside the sound horizon. This means that the consistency relation, derived for matter dominance in the previous section, can be generalized to  other background  evolutions, including $\Lambda$CDM.

For adiabatic initial conditions we can always solve eq.~\eqref{epsilon_xi_eqs} and use   eq.~\eqref{eq:lambdazeta} to relate $\epsilon$ and $\xi^i$ to the initial condition $\zeta$ and its gradient. More explicitly, we can write $\epsilon \equiv - v_L= - \hh(\eta) \zeta_L$ and $\xi^i \equiv - \vec v_L = - \hh(\eta) \partial^i \zeta_L$, where $\hh(\eta)$ is the growth function for the velocity, which one can obtain either from eq.~\eqref{epsilon_xi_eqs} or from eq.~\eqref{Euler_L} using $\epsilon = - v_L$ and $\xi^i = - \partial^i v_L$. 
%In the absence of anisotropic stress, $\delta \sigma=0$ and one simply finds $\hh(\eta)=a^{-2}(\eta)\int^\eta a^2 (\tilde \eta) \rmd \tilde \eta$. Using as initial condition in matter dominance $\zeta = - (5/3) \Phi$, yields
%If we assume that the anisotropic stress is zero, the general change of coordinates that we have to do boils down to
In terms of $\zeta_L$, the change of coordinates \eqref{etatran} and \eqref{xtran} becomes
\begin{align}
\eta &\rightarrow \eta - \hh(\eta) \zeta_L  -  \hh(\eta) x^i\partial_i \zeta_L  + 2 \hh(\eta) \partial_i \zeta_L \partial^{-2}\partial_i(\Phi_S + \Psi_S)\;, \\
x^i &\rightarrow x^i  +  \zeta_L x^i + \partial_j\zeta_L x^j x^i - \frac{1}{2} \vec x^2 \partial^i \zeta_L  - \partial^i\zeta_L \int^\eta \hh(\tilde \eta)\rmd \tilde \eta \;. \label{x_LCDM}
\end{align}
In matter dominance $\Phi_S = \Psi_S$, $\hh(\eta)=\eta/5$ and $\zeta_L = - (5/3) \Phi_L$: as expected, we recover the  transformations used in the previous subsection. Thus, apart from the different time dependence of the function $\hh(\eta)$, the form of the transformation is the same as in matter dominance. This allows us to generalize the consistency  relation straightforwardly. Since all steps in the derivation will be the same as those above, here we just summarize the main results.

{\bf Transformation of spatial coordinates}. Dilations and spatial conformal transformations only depend  on the initial condition $\zeta$ and are independent of cosmology, and thus of $\hh(\eta)$. Therefore, they
give the same result as in the case of matter dominance, i.e.,
\be
\label{conformal_lcdm}
\langle \zeta_{\vec q} \, \delta_{\vec k_1}(\eta_1) \cdots \delta_{\vec k_n}(\eta_n) \rangle'_{q\to 0} \supset - P_\zeta(q)\left( 3(n-1) + \sum_a \vec k_{a} \vec \partial_{k_{a}} + \frac 12 q^iD_i \right) \langle \delta_{\vec k_1}(\eta_1) \cdots \delta_{\vec k_n}(\eta_n) \rangle' \;.
\ee
Instead, the non-relativistic contribution coming form the time-dependent translation in eq.~\eqref{x_LCDM},
\be
\eta \rightarrow \eta \;, \qquad x^i \rightarrow x^i  -  \partial^i\zeta_L \int^\eta \hh(\tilde \eta)\rmd \tilde \eta \;,
\ee
contributes to the consistency relation as 
\be
\label{Riotto_lcdm}
\langle \zeta_{\vec q} \, \delta_{\vec k_1}(\eta_1) \cdots \delta_{\vec k_n}(\eta_n) \rangle'_{q\to 0} \supset  - P_\zeta(q) \sum_a \left( \int^{\eta_a} \hh(\eta)\rmd \eta \right) \vec q \cdot \vec k_a \langle \delta_{\vec k_1}(\eta_1) \cdots \delta_{\vec k_n}(\eta_n) \rangle' \;.
\ee

{\bf Transformation of the time coordinate.} It easy to check that for 
\be
\eta \rightarrow \eta  -  \hh(\eta) \zeta_L - \hh(\eta) x^i\partial_i \zeta_L \;, \quad x^i \rightarrow x^i \;,
\ee
the contribution to the consistency relation  from the time change of $\delta$, $\delta \to \delta + \delta \eta \cdot \delta ' $, simply generalizes to 
\be
\label{time_lcdm}
\langle \zeta_{\vec q} \, \delta_{\vec k_1}(\eta_1) \cdots \delta_{\vec k_n}(\eta_n) \rangle'_{q\to 0} \supset  -  P_\zeta(q) \sum_a \hh(\eta_a)\left( 1 + \vec q \cdot \vec \partial_{k_{a}} \right) \partial_{\eta_a} \langle \delta_{\vec k_1}(\eta_1) \cdots \delta_{\vec k_n}(\eta_n) \rangle' \;.
\ee

Now we can look at the contributions to the consistency relation coming from 
\be
\delta \to \delta + \frac{\bar \rho'}{\bar \rho}\delta\eta\; (1+\delta)  \;.
\ee
As before, we consider the contribution  from  $(\bar \rho'/ \bar \rho)\delta \eta \cdot \delta$ where $\delta \eta$ is only at first order and the one from $(\bar \rho'/\bar \rho)\delta \eta $ where $\delta \eta $ is at second order. As usual, we are not interested in higher-order terms. For the second-order $\delta \eta$ we need to use the 00 component of Einstein equations to express the short mode of the gravitational potential $\Phi_S$ in terms of $\delta$. For simplicity we assume hereafter a $\Lambda$CMD universe, in which case
$\Psi_S = \Phi_S$ and this equation reads
\be
\partial^2 \Phi_S - 3 \HH ^2 f_g (\eta)  \Phi_S = \frac{3 }{2 } \HH^2 \Omega_{\rm m} \delta \;,
\ee
where $f_g (\eta)$ is the linear growth rate of fluctuations (see e.g.~\cite{Dodelson:2003ft}) and $\Omega_{\rm m}$ is the time-dependent dark matter fractional density.
Using the equation above we then obtain
\be
\delta \to \delta + \frac{\bar \rho'}{\bar \rho}\delta\eta\; (1+\delta) = \delta +3 \mathcal H \hh(\eta)(\zeta_L + x^i\partial_i\zeta_L)\delta -18 {\cal H} \partial_i \zeta_L\frac{\mathcal H^2 \Omega_{\rm m} \hh(\eta)}{\partial^2 (\partial^2 - 3 {\cal H}^2 f_g)} \partial_i \delta  \;,
\ee
and the contribution to the squeezed limit consistency relation becomes
\be
\begin{split}
\label{background_lcdm}
\langle \zeta_{\vec q} \, \delta_{\vec k_1}\cdots \delta_{\vec k_n} \rangle'_{q\to 0} \supset &  \ 3 P_\zeta(q) \sum_a \mathcal{H}(\eta_a) \hh(\eta_a) \Bigg[  \left( 1+  \vec q \cdot \vec \partial_{k_{a}} \right)   \\
& -2 \,  \Omega_{\rm m} (\eta_a) \frac{\vec q\cdot \vec k_a}{k_a^2} \left( f_g(\eta_a) + \frac{k_a^2}{3\mathcal H^2(\eta_a)}  \right)^{-1} \Bigg]\langle \delta_{\vec k_1}(\eta_1)  \cdots \delta_{\vec k_n}(\eta_n) \rangle'  \;.
\end{split}
\ee

Summing up all contributions to the consistency relation we finally get
\begin{align}
&\langle \zeta_{\vec q} \, \delta_{\vec k_1}(\eta_1) \cdots \delta_{\vec k_n}(\eta_n) \rangle'_{q\to 0} =  -  P_\zeta(q)\left[ 3(n-1) + \sum_a \vec k_{a} \cdot \vec \partial_{k_{a}} + \sum_a \hh(\eta_a)(\partial_{\eta_a}-3\mathcal H(\eta_a))  \right. \nonumber \\
& \quad \qquad  \left.  +\sum_a \left( \int^{\eta_a} \hh(\eta)\rmd \eta \right) \vec q \cdot \vec k_a +\frac 12 q^iD_i  + \sum_a \hh(\eta_a)(\partial_{\eta_a}-3\mathcal H(\eta_a)) \vec q \cdot \vec \partial_{k_{a}} \right. \nonumber \\
&  \quad \qquad \left. + 6 \sum_a \Omega_{\rm m} (\eta_a) \hh(\eta_a)\mathcal H(\eta_a)\frac{\vec q\cdot \vec k_a}{k_a^2} \left( f_g(\eta_a) + \frac{k_a^2}{3\mathcal H^2(\eta_a)}  \right)^{-1} \right] \langle \delta_{\vec k_1}(\eta_1) \cdots \delta_{\vec k_n}(\eta_n) \rangle' \;.
\end{align}
This equation is derived under assumptions that the anisotropic stress is zero but this can be easily relaxed.

Let us consider the non-relativistic limit  of this expression. As in matter dominance, the first term in the second line dominates the consistency relation. Moreover, if for the long wavelength mode we define the growth function of the density field $D_\delta $ by $\delta_L = D_\delta (\eta) \nabla^2 \zeta_L$, from the continuity equation $\delta_L' + \vec \nabla \cdot \vec v_L =0$ we find $D_\delta (\eta) = - \int^\eta D_v (\tilde \eta) \rmd \tilde \eta$ and the above expression reduces to
\be
\label{NRgeneral}
\langle \delta_{\vec q} (\eta) \, \delta_{\vec k_1}(\eta_1) \cdots \delta_{\vec k_n}(\eta_n) \rangle'_{q\to 0} =  - P_\delta(q,\eta) \sum_a \frac{D_\delta(\eta_a)}{D_\delta(\eta)} 
\frac{\vec q \cdot \vec k_a}{q^2} \langle \delta_{\vec k_1}(\eta_1) \cdots \delta_{\vec k_n}(\eta_n) \rangle'  \;.
\ee

\section{Conclusions and future directions}
In single-field models of inflation, a long mode that has been outside the sound horizon since inflation has no physical effect on short-scale modes and it boils down to a redefinition of the spatial and time coordinates. In this paper we used this idea to write consistency relations for the late universe described by CDM or $\Lambda$CDM. The generalization to dark energy models with negligible fluctuations or with very small speed of sound should be straightforward.  
These relations do not assume that the short modes can be treated perturbatively, and as such they are valid also when the short modes are deep in the non-linear regime, even when we take into account the effect of baryons and their bias. Indeed it is important to notice that even if we wrote our relation thinking about $\delta$ as the dark matter density contrast, nothing would change if instead we used $\delta_h$ or $\delta_g$, i.e.~the number density of haloes or of galaxies. Since the long mode does not affect the physics at short scales, it will not modify in any way the (linear or non-linear) bias. 
%{\bf Neutrinos}

Many generalizations of this technique are possible. One can write consistency relations involving short scale velocities, take the long mode to be a tensor and not a scalar mode, consider correlation functions with more than one long mode or with soft internal lines, exactly as one does for the primordial correlation functions in single-field inflation.

Of course our consistency relations are not directly observable, since one has to take into account all the Newtonian and relativistic effects which relate the correlation functions of the density contrast and the velocity with the actual measurement of the galaxy position and redshift (see for example \cite{Yoo:2009au,Bonvin:2011bg,Challinor:2011bk,Yoo:2012se}). The next natural step would be to write consistency relations for the observed quantities, in the same way one does it in the case of the CMB bispectrum \cite{Creminelli:2004pv,Creminelli:2011sq}.

Our relations encode the primordial consistency relations of single-field inflation and as such can be considered as a test of this minimal cosmological scenario. Conversely any deviation will robustly rule out any single-field model, similarly to what happens for the scale-dependent bias first discussed in \cite{Dalal:2007cu}. 
For example one could consider the squeezed 3-point function with a long mode and two arbitrarily short modes of galaxy density. Our relations state that this 3-point function is only due to ``projection effects'' as a consequence of the change of coordinates induced by the long mode. In the presence of primordial local non-Gaussianity we expect larger effects due to the local change of the power spectrum, that is modulated by the long mode. This effect may be difficult to estimate, both because the short modes may be in the non-linear regime and because one has to deal with the physics of galaxy formation. Still this may not be so problematic: any effect that deviate from our consistency relation would be a sign of multifield dynamics. This kind of approach is worth further studies.

Besides the absence of primordial non-Gaussianity, another hypothesis of our relations is the equivalence principle. Since the equality of gravitational and inertial mass is violated in models of modified gravity, for example in the presence of chameleon screening \cite{Hui:2009kc}, one can look for deviations of the consistency relations above as a smoking gun of violations of the equivalence principle on large scales. Let us for instance focus on the non-relativistic limit, eq.~\eqref{NRgeneral}, and take the 3-point function of a long mode and two short modes of the number density of two {\em different} kind of objects, suspected to have a different ratio of gravitational and inertial mass. The equal time correlator should vanish at order $1/q$ if the effect of the long mode is a common acceleration of the two populations, this statement being independent of the biases of the two populations and of the growth functions. A residual divergence $1/q$
as $q \to 0$ would show a violation of the equivalence principle. We will come back to this topic in a future publication. 

Notice that the assumptions of our relations---single-field inflation and Equivalence Principle---are two sides of the same coin. They are consequence of diffeomorphism invariance and they are violated when more degrees of freedom become relevant, either during inflation or in the late universe. From this point of view, they can be regarded as ``Single-Field Consistency Relations'' in a generalized sense.

\subsection*{Acknowledgements}
It is a pleasure to thank F.~Bernardeau, E.~Castorina, B.~Horn, L.~Hui, A.~Paranjape, M.~Pietroni, A.~Riotto, E.~Sefusatti, L.~Senatore, R.~Sheth and X.~Xiao for useful discussions. In particular we thank M.~Pietroni for useful correspondence about models that violate the Equivalence Principle. We acknowledge work in progress both by L.~Hui, B.~Horn and X.~Xiao and by A.~Kehagias and T.~Riotto on related issues. JN is supported by ERC grant FP7-IDEAS-Phys.LSS 240117. FV acknowledges partial support by the ANR {\it Chaire d'excellence} CMBsecond ANR-09-CEXC-004-01.

\appendix
\section{Galilean or not Galilean?}
\label{sec:Galilean}
In this appendix we want to show that our construction of the adiabatic mode and the consequent consistency relations are {\em not} a consequence of Galilean invariance, not even in the non-relativistic, sub-Hubble regime, but of the equivalence principle. Similarly, the cancellation of the effect of long modes in perturbation theory for equal-time correlators \cite{Jain:1995kx,Scoccimarro:1995if,Bernardeau:2011vy,Bernardeau:2012aq,Blas:2013bpa,Carrasco:2013sva} is {\em not} a consequence of Galilean invariance, but again of the equivalence principle, as recently pointed out in \cite{Carrasco:2013sva}. Therefore the adjective ``Galilean'' sometimes used in the literature \cite{Scoccimarro:1995if,Peloso:2013zw} should be avoided. 

In the non-relativistic limit, the equations that govern the motion of a perfect fluid in the presence of gravity are given by
\begin{align}
\label{fluid_gravity}
& \dot \rho + \nabla (\rho \vec v) = 0 \;, \nonumber \\
& \dot{\vec v} + (\vec v \cdot \vec \nabla)\vec v = - \frac{\nabla p}{\rho} - \nabla \phi \;, \nonumber \\
& \nabla^2 \phi = 4\pi G \rho \;.
\end{align}
These equations are invariant under a Galilean transformation
\be
\label{galilean_transf}
\vec x' = \vec x - \vec v_0 t\;, \quad t'=t \;.
\ee
The presence of the fluid breaks the Galilean symmetry spontaneously: there is a preferred reference frame in which the fluid is at rest. A non-linearly realized symmetry says something about the dynamics of the long-wavelength excitations of the systems, i.e.~the Goldstone bosons. Naively---and incorrectly!---one would say that the adiabatic mode we have studied is the Goldstone boson associated with Galilean invariance. Let us see that this is not the case.

For simplicity, let us start with gravity turned off.  If we do a Galilean boost, we generate a mode that has constant velocity $\vec v_0$ and unperturbed density, $\delta=0$. This is a physical mode, i.e.~it is a solution in the long wavelength limit, $q \to 0$, of the linearized equations of motion
\be
\dot \delta + \vec \nabla \cdot \vec v = 0\;, \qquad \dot{\vec v} = - \frac{\nabla p}{\rho} \;.
\ee
Now let us add gravity---and set the pressure to zero for simplicity---and describe the FRW solution in physical coordinates. The equations of motion \eqref{fluid_gravity} with $p=0$ are solved by 
\be
\vec v = H \vec x \;, \qquad \rho = 3M_{\mathrm{Pl}}^2 H^2 \;, \qquad \phi=\frac 14 H^2 x^2  \;, \qquad H=\frac{2}{3t} \;.
\ee
If we now do a Galilean transformation \eqref{galilean_transf} at first order in $v_0$ we generate a new solution, 
\be
\vec v = H \vec x - \frac 13 \vec v_0 \;, \qquad \rho = 3M_{\mathrm{Pl}}^2 H^2 \;, \qquad \phi=\frac 14 H^2 x^2 + \frac 12 H^2 \vec v_0 \cdot \vec x \; t \;,
\ee
where, for simplicity, we have dropped the primes.
Indeed, it is easy to check that the new $\rho$, $\phi$ and $\vec v$ satisfy the equations of motion. 
Notice that the generated perturbation of the gravitational potential has the following time dependence
\be
\delta\phi \propto \frac 1t \;.
\ee

The same result can be obtained if we write the standard perturbation theory in comoving coordinates and conformal time. The equations of motion for perturbations have the following form, 
\begin{align}
& \delta' + \nabla [(1+\delta) \vec v] = 0 \;, \nonumber \\
& \vec v ' + (\vec v \cdot \vec \nabla)\vec v + \mathcal H \vec v = - \nabla \Phi \;, \nonumber \\
& \nabla^2 \Phi = \frac 32 \mathcal H^2 \delta \;,
\end{align}
where $\mathcal H = 2/\eta$. After some algebra, it can be shown that in these coordinates a Galilean transformation has the following form \cite{Scoccimarro:1995if}
\be
\label{eq:Galileo}
\vec x'=\vec x -\frac 13 \eta \,\vec v_0 \;, \qquad \vec v ' = \vec v - \frac 13 \vec v_0 \;,
\ee
and, given that the total Newtonian potential is $\phi = \frac 14 \mathcal H^2 x^2 + \Phi$, we also induce a homogeneous gradient of $\Phi$ with the same time dependence discussed before (taking into account the difference between physical and comoving coordinates), i.e.,
\be
\label{eq:Goldbehave}
\delta\Phi = \frac 1 3 \mathcal H \vec v_0 \cdot \vec x \propto \frac 1\eta \;.
\ee
As in the case without gravity, we would expect this mode to be the long wavelength limit of a physical solution---and maybe to be related to the adiabatic mode discussed in the main text---but this is {\em not} the case. Indeed the linearized equation for $\delta$,
\be
\delta'' + \mathcal H \delta ' - \frac 32 \mathcal H^2 \delta = 0\;,
\ee
gives the well-known solutions
\be
\Phi \propto \mbox{const.}  \quad \mbox{and} \quad \Phi \propto \eta^{-5} \;.
\ee 
Thus, the time-dependence \eqref{eq:Goldbehave} of the Goldstone is not a possibile solution! Formally, for $q=0$, the Goldstone solution exists but it cannot be extended to nonzero momentum. 

The reason why this happens is that gravity is a long range interaction and this makes the $q \to 0$ limit singular: indeed $\Phi$ is a non-local function of $\delta$. In other words, gravity is gauging the Lorentz (and thus Galileo) group and once a symmetry is gauged we do not expect the Goldstone mode to survive: it will be ``eaten'' by the graviton and get a mass of order $H$, similarly to what happens in the Higgs mechanism.\footnote{Of course, at short scales the Goldstone mode decouples and we recover the case without gravity, but this requires to concentrate on time scales much shorter than $H^{-1}$.}

A posteriori it is easy to see that Galilean invariance is a red herring. Indeed, the change of coordinates in the non-relativistic limit \cite{Kehagias:2013yd,Peloso:2013zw} that generates our adiabatic solution---see eq.~\eqref{eq:changec}--- is
\be
\eta \rightarrow \eta \;,\qquad x^i \rightarrow x^i + \frac{1}{6}\eta^2 \partial_i\Phi_L \;,
\ee
which is {\em not} a Galilean transformation, eq.~\eqref{eq:Galileo}. 
We are adding a time-dependent (we have a time-independent gradient in comoving coordinates, in physical coordinates it depends on time), homogeneous gravitational force via a change of coordinates, which corresponds to an homogeneous acceleration: the prototypical example of the equivalence principle.\footnote{This mode, indeed, solves the equations of motion. The induced gravitational potential and velocity are given by
\be
\Phi = x^i \partial_i \Phi_L \;, \qquad v_i=  -\frac{1}{3} \eta \,\partial_i\Phi_L \;,
\ee
and it is easy to check that continuity and Euler equations are satisfied. Moreover, the gravitational potential $\Phi$ is time independent: $\Phi$ is a good $q\to 0$ limit of a physical mode.}
To make the distinction even clearer, one can consider adding a test particle that violates the equivalence principle, i.e.~with a different ratio of gravitational and inertial mass. In this case Galilean invariance is obviously untouched but the change of coordinates does not generate a physical solution, because the test particle has a different acceleration with respect to the others as a consequence of the violation of the equivalence principle. In this case, the long mode would be coupled to short modes not only through the tidal forces and this would impair the discussed cancellation for the equal-time correlators, which has thus nothing to do with Galilean invariance.

\end{document}